\documentstyle[preprint,prc,aps,epsf,amssymb]{revtex}

\newcommand{\bmq}{{\mbox{\boldmath $q$}}}

\begin{document}
\preprint {WIS-02/15 Apr-DPP}
\draft
\date{\today}
\title{Extraction of the neutron structure function $F_2^n$ from 
inclusive scattering data on composite nuclei}
\author{A.S. Rinat and M. F. Taragin}
\address{Weizmann Institute of Science, Department of Particle Physics,
Rehovot 76100, Israel}
\maketitle
\begin{abstract}

We consider a generalized convolution, linking Structure Functions (SF)
$F^N_2$ for nucleons, $F^A_2$ for a physical nucleus and  $f^{PN,A}$ for a  
nucleus, composed of point-nucleons. In order to extract $F_2^n$ we employ 
data on $F_2^{p,A}$ and the computed $f^{PN,A}$. 
Only for $Q^2\approx 3.5\,{\rm GeV}^2$ 
do data permit the extraction of $F_2^A(x,3.5)$ over a sufficiently
wide $x$-range. Applying Mellin transforms, the above relation between SF 
turns into an algebraic one, which one solves for the Mellin transform of the 
unknown $F_2^n$. We present inversion methods leading to the desired $F_2^n$, 
all using a parametrization for $C(x,Q^2)=F_2^n(x,Q^2)/F_2^p(x,Q^2)$. Imposing 
motivated constraints, the simplest parametrization leaves one free parameter 
$C(x=1,Q^2)$. For $Q^2= 3.5\,{\rm GeV}^2$ its average over several targets 
and different methods is $\langle C(1,3.5)\rangle=0.54\pm0.03$. We argue 
that for the investigated $Q^2$, 
$C(x\to 1,3.5)$ is determined by the nucleon-elastic ($NE$) part of  SF. A 
calculation of the latter comes close to the extracted value. Both are close
to the SU(6) limit $u_V(x,3.5)=2d_V(x,3.5)$ for parton distribution functions.                   

\end{abstract}

\bigskip

It has been recently proposed to extract the ratio $C(x,Q^2)=
F_2^n(x,Q^2)/F_2^p(x,Q^2)$ of the $n,p$  structure functions (SF) from data 
on ratios $\sigma(^3{\rm H})/\sigma(^3{\rm He})$  of inclusive cross sections. 
An electron beam with $E\approx$ 11-12 GeV \cite{petr} is expected to provide
the required  wide, continuous kinematic ranges.  The proposal has already 
elicited discussions, mainly on the stability of the suggested 
extraction techniques for $C(x,Q^2)$ \cite{afn,saito,pace,sss}. 
 
An experiment, involving the SF of the lightest isobars  with minimal $I$-
spin symmetry breaking is presumably best suited to obtain $C(x,Q^2)$. 
However, even when an upgrading of the beam will be approved, results on the 
$A$=3 proposal are not expected before 2006. One therefore wonders whether existing 
data on $F_2^A$ may furnish similar information. Below we explore 
this possibility and present first results for $C(x,Q^2)$ and $F_2^n(x,Q^2)$
from inclusive scattering data on D,C and Fe.  

In the following we limit ourselves to inclusive scattering of unpolarized 
electrons from randomly oriented targets $A$. The cross section per 
nucleon for beam energy $E$ reads
\begin{eqnarray}
\frac{d^2\sigma^{eA}(E;\theta,\nu)/A}{d\Omega\,d\nu}
=\frac{2}{M}
\sigma_M(E;\theta,\nu)F_2^A(x,Q^2)\bigg\lbrack\frac {xM^2}{Q^2}+
{\rm tan}^2(\theta/2)\frac {F_1^A(x,Q^2)}{F_2^A(x,Q^2)}\bigg\rbrack\ ,
\label{a1}
\end{eqnarray}
where  $\theta$  and $\nu$ are  the scattering angle and the energy loss.
$F_{1,2}^A(x,Q^2)$ are  nuclear structure functions (SF) per nucleon, 
expressed in terms of the squared 4-momentum transfer $Q^2=\bmq^2-\nu^2$ 
and the Bjorken variable  $x=Q^2/2M\nu$ ($M$ is the nucleon mass)
with range $0\le x \le A$. 

Henceforth nuclear and nucleon SF are assumed to be related by some 
generalized convolution  $F^A=f*F^N$, for instance \cite{aku,gr1}
\begin{mathletters}
\label{a2} 
\begin{eqnarray}
F^A_k(x,Q^2)&=&\int_x^A\frac {dz}{z^{2-k}} \bigg \lbrack \frac{Z}{A}
f_p^{PN,A}(z,Q^2) F_k^p(x/z,Q^2)+\frac{N}{A}f_n^{PN,A}(z,Q^2)
F_k^n(x/z,Q^2)\bigg\rbrack
\label{a2a}\\
&\approx& \int_x^A\frac {dz}{z^{2-k}} f^{PN,A}(z,Q^2)
F_k^{\langle N\rangle}(x/z,Q^2)\ .
\label{a2b}
\end{eqnarray}
\end{mathletters}
In Eq. (\ref{a2}) $f_{p,n}^{PN,A}$ is the SF for a nucleus, composed of 
point-nucleons, where initially a $p$ or a $n$ absorbs the virtual photon; 
$f^{PN,A}$ is their average over the number of protons and neutrons in the 
target $A(Z,N)$. $F_k^{\langle N\rangle}$ stands for the similarly averaged  
nucleon ($N$) SF (We often drop arguments of functions when there is no 
danger of confusion) \footnote {* In Eqs. (\ref{a2}) appear admixtures of 
nucleon SF \cite{atw}, which are negligible for the involved $Q^2$.}. 
\begin{eqnarray}  
f^{PN,A}&=&\frac{Z}{A}f_p^{PN,A}+\frac {N}{A}f_n^{PN,A}
\nonumber\\
F_k^{\langle N \rangle}&=&\frac{Z}{A}F_k^p+\frac {N}{A} F_k^n
\label{a3}        
\end{eqnarray}  
Eq. (\ref{a2}) describes partons from nucleons in a nucleus, but not those
from other  sources (virtual bosons) neither does it account for
anti-screening.  Both limit the use of (\ref{a2}) to $x\gtrsim$ 
0.15-0.20 \cite{pion}. Eq. (\ref{a2}) is estimated to be valid for  
$Q^2\gtrsim Q_c^2\approx 2-2.5$ GeV$^2$ \cite{rt3,commar}.

We briefly mention previously suggested extraction methods for  $F_2^n$. 
Those dealt mostly with D data and use $f^{PN}$ in the Impulse 
Approximation (IA) \cite{bod1,liut}. It has for instance been emphasized 
that Eq. (\ref{a2}) is a Fredholm integral equation for the unknown 
$F_k^{\langle N\rangle}$. Discretization in $x$ and $z$ produces a set of 
linear equations with a solution, tending to the exact answer for $\Delta z
\to 0 $ \cite{umni}. The strong variation of the $'$kernel$'$ $f^{PN}$  
apparently hampers an actual application of the above. We also mention 
iteration 
methods to deconvolute nuclear effects for a D target \cite{bod1,meln}. 

A different deconvolution has been suggested for the proposed precision
data, 
relating $F_k^A$ \cite{petr}. For those one may define super-ratios 
\cite{afn,pace}
\begin{mathletters}                      
\label{a4}                                   
\begin{eqnarray}                           
{\cal R}^{A_1,A_2}(x)&\equiv&\frac {\rho^{A_1}(x)}{\rho^{A_2}(x)}
\label{a4a}\\                            
\rho^A(x)&=& \frac{[F_2^A(x)/F_2^p(x)]}{a[F_2^n(x)/F_2^p(x)]+b}\,\,\,,
\label{a4b}
\end{eqnarray}
\end{mathletters}
with, in principle, arbitrary  $a$ and $b$. For given $F_2^p$ those ratios 
contain $C(x)$ explicitly, and are from Eq. (\ref{a2}) seen to depend  
implicitly on $C(x/z)$. Iteration determines $C(x)$.

Finally we recall the expansion of $F^{\langle N\rangle}(x/z)$ in Eq.
(\ref{a2}) around the maximum  $z\approx1$ of $f(z)$  \cite{akul}
\begin{eqnarray}
F_2^A(x)&=&F_2^{\langle N\rangle}(x)+{\cal M}_1(x)
[xF_2^{\langle N\rangle(1)}(x)]+{\cal M}_2(x) [xF_2^{\langle N
\rangle(1)}(x)
+\frac{x^2}{2}F_2^{\langle N\rangle(2)}(x)]+.... \ ,
\label{a5}
\end{eqnarray}
with ${\cal M}_n(Q^2)=\int_0^A dz f^{PN,A}(z,Q^2)(1-z)^n$. The first 
$A$-independent term $F_2^{\langle N \rangle}(x)$
adequately represents the series for $x\lesssim 0.4$.
Insufficient knowledge of $F_2^{\langle N\rangle(l)}(x)$, the $l$-th 
derivative of $F_2^{\langle N\rangle}(x)$, prevents the use of the expansion 
(\ref{a5}) beyond that range. Although not the region of prime interest, the
low-$x$ result will guide the extraction of $F_2^n$.

We first recall calculations of  $F_k^A$ from Eq. (\ref{a2}) with $F_k^{p,n}$ 
as input. For nucleons in vacuum there are abundant data on $F_2^p$ 
\cite{amad} and less accurate ones for $F_1^p$ \cite{bod}. Inclusive data do  
not reach the elastic point $x=1$, but the used parametrizations do.
With no direct information on 
$F_k^n$, one frequently uses the 'primitive' assumption
\begin{eqnarray}
F_k^{n,pr}\equiv 2F_k^D-F_k^p, 
\label{a6}
\end{eqnarray}
which is the first term in Eq. (\ref{a5}) for the D
and thus holds only for small $x$. The question of interest 
is  how modifications of Eq. (\ref{a6}) for $x\gtrsim 0.4$ influence 
$F_k^A(x,Q^2)$, Eq. (\ref{a2}). 

The  SF $f^{PN,A}$ is a many-body property and requires a model for a 
calculation, such as the perturbative IA, which to lowest order contains the 
single-hole spectral function \cite{cio,om1}. There nucleons appear off their 
mass shell and a prescription is needed to handle those. We  prefer a 
non-perturbative version, based on a generalized Gersch-Rodriguez-Smith 
(GRS) theory, where the SF of nucleons $F^N$ are reasoned to be on 
their mass shell. That method moreover allows a computation of $f^{PN,A}$ 
beyond its lowest order \cite{gr1,commar,rt1,gr2,rt11}. 

The above program has initially been realized for $A\ge 12$. It has been 
shown that $f^{PN,A}$ is nearly independent of $A$, and using Eq. (\ref{a2}) 
one proves the same for $F_k^A$. Data indeed show that for $x\lesssim 0.8$, 
the ratios $\mu^{A,A'}=[d^2\sigma^{eA}/A]/[d^2\sigma^{eA'}/A']\approx 
F_2^{A}/F_2^{A'},\,\, A,A'\ge 12$ differ differ from 1 by less than 
2-3$\%$ \cite{rt1,rt11,arrold}.

The above does not hold for the lightest nuclei, in particular not for 
$A'\to D$, when
\begin{eqnarray}
\mu^A(x,Q^2)=\mu^{A,D}(x,Q^2)=\frac {d^2\sigma^{eA}(x,Q^2)/A}
{d^2\sigma^{eD}(x,Q^2)/2} \approx \frac {F_2^A(x,Q^2)}{F_2^D(x,Q^2)} 
\label{a7}
\end{eqnarray}
are the  EMC ratios, with their characteristic deviations from 1 in the range 
$x\lesssim 0.9$ \cite{gomez}. 

For $A\le 4$ and  given $NN$ interaction, one may presently compute with 
great precision nuclear ground states \cite{bench}, as well as $f^{PN,A}$ in 
Eq. (\ref{a2}). Using those, GRS calculations of inclusive cross sections 
have recently been completed for $D$ \cite{rtd} and $^4$He \cite{viv}.

We now address the inverse problem of trying to extract $F_2^n$ from data on 
$F_2^{p,A}$ and the above-mentioned computed $f$. First one needs to obtain
nuclear SF and we shall use JLab data for D \cite{nicu1,arrd} and for 
C, Fe \cite{arr1}, supplemented by some older NE3 data of more restricted 
kinematics \cite{ne3}. We mention two approaches: 

i) $R$-ratios:  Ideally one performs a Rosenbluth extraction of $F_k^A$ 
from cross sections (\ref{a1}) for fixed $x,Q^2$ and varying $\theta$ 
or $E$, which provides
\begin{eqnarray}
R^A(x,Q^2)=\bigg (1+\frac{4M^2x^2}{Q^2}\bigg)
\bigg (\frac{F_2^A(x,Q^2)}{2xF_1^A(x,Q^2)}\bigg )-1
\label{a8}
\end{eqnarray}
Unfortunately, the set of $(x,Q^2)$ points in the JLab and NE3 data does not 
enable the above Rosenbluth extractions. Instead one frequently invokes an 
empirical expression such as $R^A(x,Q^2)\approx 0.32/Q^2$. It assumes $A$, as 
well as $x$-independence and prescribes the dependence on $Q^2$ \cite{dasu}.
The above empirical $R$ actually contradicts theoretical
predictions \cite{commar}. We shall return to this point below.
 
ii) $Theoretical\, information$: Using the primitive $F^{n,pr}$, computed 
cross sections (\ref{a1}) as function of $\theta, \nu$ generally agree with 
data in the deep-inelastic regions, $x \lesssim 1,\,(Q^2\gtrsim Q_c^2$). 
There we select data, for which the relative deviation $\alpha (=
\alpha(x,Q^2))=(F_2^{A,th}-F_2^{A,exp})/F_2^{A,exp}$ satisfies 
$|1-\alpha|\lesssim 0.2$ and which, for given $\theta$  appear to change 
smoothly with $x$. We suggest to attribute the above deviations in equal 
measure to both structure functions $F_k^A$ in (\ref{a1}), which enables the
definition of quasi-data 
\begin{eqnarray}
F_k^{A,qd}(x,Q^2) \equiv \alpha(x,Q^2) F_k^{A,th}(x,Q^2)  
\label{a9}
\end{eqnarray}
By construction, the $F_k^{A,qd}$ produce an exact fit to cross sections. 

Use of Eq. (\ref{a2}) for the purpose of inversion requires input for fixed 
$Q^2$ and running $x$, whereas cross section data are for essentially one 
beam energy $E$ and cover 
separated $x$ bins. $Q^2$ varies mildly within those $x$-bins, but 
not when going from one bin to a neighbouring one. It is thus necessary to 
make cuts for fixed $Q^2$, containing a maximal number of $x$-points with 
$x\lesssim 1.0-1.1$. After careful interpolation between quasi-data, the 
above appears only fulfilled for $Q^2$ between 3-4 GeV$^2$ and then only for 
$0.5 \lesssim x\lesssim 1.1$. It is clearly impossible to reliably
extrapolate the manipulated quasi-data into the crucial range $x\lesssim 0.5$. 

At this point we invoke the empirical observation that for 
$x_0\approx 0.16-0.18\,$, $F_2^p(x_0,Q^2)$ and $F_2^D(x_0,Q^2)$ are 
practically constant as function of $Q^2$ \cite{amad}. We exploit this by
considering Eq. (\ref{a2}) for small $x\le x_0$. Upon substitution of the 
small-$x$ part of Eq. (\ref{a5}) into (\ref{a2}) and use of 
$\int_0^A\,dz f^{PN,A}(z)=1$, one easily shows that for
$any$ target 
\begin{mathletters}
\begin{eqnarray}
\label{a10} 
F_2^{A, nucl}(0.18)&\approx &F_2^N(0.18)\approx 0.34 
\label{a10a}\\
F_2^{A, nucl}(0)&=&F_2^{\langle N\rangle}(0)=F_2^N(0)\ne 0
\label{a10b}
\end{eqnarray}
\end{mathletters} 
The latter extension to  $x=0$ is in agreement with Eq. ({\ref{a5}) and will
be needed below  \footnote{* The above explains why for all $A$ and $Q^2$, 
EMC ratios $\mu^A(x,Q^2)$ cross 1 for $x\approx 0.18$\cite{gomez}.}.

We recall that  $f^{PN,A}$ in Eq. (\ref{a2}) relates only to nucleons in the 
target and as a result, Eqs. (\ref{a10}) hold only 
for the nucleonic component $F_2^{A,nucl}\,\,$. 
That component may now be $interpolated$ for $x \lesssim 0.55$, completing 
knowledge of $F_2^{A,nucl}(x)$ over the entire relevant range 
$x\lesssim 1.1$. 

After the critical remarks on the empirical form of the $R$-ratio it comes
as a surprise that in the range $0.5 \lesssim x\lesssim 1$, 
$F_2^{Fe,qd}(x,3.5)$ and the empirically extracted $F_2^{Fe(R)}(x,3.5)$ agree 
to within $\pm  5\%$, and even for $1\lesssim x \lesssim 1.5$ to within 
$\pm 12\%$!  It vindicates
Arrington's claim that even a 100$\%$ uncertainty in the phenomenological  
$R$ incurs only a 5$\%$ uncertainty in $F_2^{A(R)}$ \cite{arrd}.

We proceed as follows. For Fe, for which the number of available 
deep-inelastic data points is largest, we use $F_2^{Fe,qd}$ from
method ii). With insufficient $C,D$ data for application of that 
method, we exploit Eq. (\ref{a7}) and find 
\begin{eqnarray}
F_2^{C,qd}&\equiv&\frac {F_2^{C,R}}{F_2^{Fe,R}}F_2^{Fe,qd} 
\nonumber\\
F_2^{D,qd}&=&[\mu^C]^{-1}F_2^{C,qd}
\label{a11}
\end{eqnarray}
EMC data for $\mu^C$ are in the desired range $0.3\lesssim x\lesssim 0.80$
where they require some smoothing. For $x\lesssim 0.30$ we interpolated the 
purely nucleonic component as in ii) and added for $0.80<x<0.88$  
averages of Be and Al data \cite{gomez}.

We now introduce Mellin transforms (MT) and their inverses, which for real 
$g(x)$ are defined as
\begin{mathletters}
\label{a12}
\begin{eqnarray}
\tilde g(u)&=&\int_0^{\infty} dx x^{u-1} g(x)
\label{a12a}\\
g(x)&=&\frac {1}{\pi}\int_0^{\infty} dt\, {\rm Re}\bigg \lbrack
\frac {g(a+it)}{x^{a+it}}\bigg \rbrack,
\label{a12b}
\end{eqnarray}
\end{mathletters}
The constant $a$ is chosen such, that $g(u)=g(a+it)$ is free of singularities 
in the complex $u$ plane to the right of the imaginary $u$ axis shifted by $a$. 

Application of Eq. (\ref{a12a}) to  Eq. (\ref{a2}), turns the generalized 
convolution into a linear relation, which can be solved for 
$\tilde F_2^n(u,Q^2)$. For fixed $Q^2$
\begin{mathletters}
\label{a13}
\begin{eqnarray}
\tilde F_2^A(u)&=&\tilde f^{PN,A}(u+1) \tilde F_2^{\langle N\rangle}(u)
+{\cal G}^{\langle N(NE)\rangle} 
\label{a13a}\\
\tilde {\cal F}_2^{n(A)}(u)&=&\frac {A}{N}
\frac {\tilde F_2^A(u)}{\tilde f^{PN,A}(u+1)}- \frac {Z}{N}\tilde F_2^p(u)    
-\bigg [\frac {Z}{N}{\cal G}^{p}+{\cal G}^{n} \bigg ],
\label{a13b}
\end{eqnarray}
\end{mathletters}
with ${\cal G}^{N,NE}$, 
combinations of the standard nucleon static formfactors, making up the $NE$
parts of $F_2^{N,NE}$ (see for instance 
Ref. \onlinecite{commar}). Above we denote by $\tilde {\cal F}_2^{n(A)}$ the 
$A(N,Z)$-dependent right-hand side of  Eq. (\ref{a13b}). For exact input 
$F_2^A$, it should coincide with $\tilde F_2^n(u)$, the MT of $F_2^n(x)$. 

Next we consider $u=0$ in Eq. (\ref{a13b}). Since the 
normalization of $f^{PN,A}$ implies $\tilde f^{PN,A}(1)=1$, one finds
\begin{eqnarray}
\tilde {\cal F}_2^{n(A)}(u=0)=\frac {A}{N} \tilde F_2^{A,nucl}(u=0)-
\frac{Z}{N}\tilde F_2^p(u=0)-\bigg [\frac {Z}{N}{\cal G}^{p}+{\cal
G}^{n}\bigg ]   
\label{a14}
\end{eqnarray}
Consequently, Eq. (\ref{a12a}), when used in (\ref{a10}) implies that 
none of the MT of  SF are defined for $u=0$. This has numerical consequences
also in the immediate neighbourhood of $u=0$.

Unfortunately, the direct inversion Eq. (\ref{a12b}) of $\tilde {\cal F}(u)$ 
runs into serious numerical problems. We therefore take recourse to indirect 
methods, all featuring a parametrization of the ratio $C$ in
\begin{eqnarray}
F_2^n(x,Q^2)=F_2^n(x,Q^2;d_k)&=&C(x,Q^2;d_k)F_2^p(x,Q^2)
\nonumber\\
C(x,Q^2;d_k)&=&\sum_{k\ge 0} d_k(Q^2)(1-x)^k
\label{a15}
\end{eqnarray}
Eq. (\ref{a10}) implies a first constraint $C(0)(=F_2^n(0)/F_2^p(0))=
\sum_{k\ge 0}d_k(Q^2)=1$  
Next, one may exploit $F_2^n(x)=F_2^{n,pr}(x)$, Eq. (\ref{a5}), for small 
$x\lesssim 0.35$. We use only $F_2^n(0.2,3.5)=0.75$ which for 
$k_{max}=2$ leaves one free parameter $d_0=C(1)$.

It is convenient to re-parametrize  $F_2^p(x,3.5)$ as follows 
\begin{mathletters}
\begin{eqnarray}
\label{a17}
F_2^p(x,Q^2)&=x^{-a^2}&\sum_{m\ge 1} c_m(1-x)^m ;x\ge 0.02
\label{a17a}\\
&=&0.42~~~~~~~~~~~~~~~~;x\le 0.02     
\label{a17b}
\end{eqnarray}
\end{mathletters}
A small $a^2\ll 1$ in Eq. (\ref{a17a}) produces the observed rise of 
$F_2^p(x)$  for small $x$, while the cut-off in Eq. (\ref{a17b}) avoids the 
singularity in
(\ref{a17a}). Eq. (\ref{a10}) again implies the same for $F_2^{A,n}(x)$.
In the region $0.02 \lesssim x\lesssim 0.9$, $F_2^p$, Eq. (\ref{a17a}), 
practically coincides with the parametrization in Ref. \onlinecite{amad}, 
both agreeing with data averaged over resonances. 

We now explore some extraction methods, comparing relevant computed 
or extracted expressions $A$, related to $F_2^n$ and parametrized forms $B$:

I) The extracted $\tilde {\cal F}_2^{n(A)}(u)$, 
Eq. (\ref{a13b}) and the  MT of $F_2^n$, Eq. (\ref{a15}).
 
II) $F_2^{A,qd}(x)$, Eq. (\ref{a9}), with the nuclear SF 
$F_2^A(x)$, computed from (\ref{a2}), now using Eq. (\ref{a15}).

The parameters $d_k$, are determined by minimization of variances 
$w(d_0)=\sum_i\bigg |A_i-B_i \bigg |^2$ (or of relative variances). 
Table I summarizes our results. The spread  of the individual entries $C(1)$ 
from I) reflects  variations in the smallest and largest $u$, 
retained in the above sum (see remark after Eq. (\ref{a14})). 
An empty entry indicates the absence of a minimum in the studied 
intervals. We note that in those cases the slope of the variance is very small 
and that the range of $C(1)$ is actually compatible with values from real 
minima. The results for the two methods and for the three targets produces 
a well-determined average $\langle C(1,3.5)\rangle=0.54\pm0.03$.

In Fig. 1 we show the full $C(x,3.5)$, as well as  $F_2^{p,n}(x,3.5)$. On 
the right abscissa are marked the values 2/3, 3/7, 1/4 for $C(1)$, 
corresponding 
to exact SU(6) symmetry, dominance of $S=0$ over $S=1$ di-quark coupling 
and the same for the $z$-component $S_z$ for $x\to 1$ (see for instance Ref. 
\onlinecite{meln})
\footnote {* The extracted $F_2^n$ allows an evaluation of the Gottfried 
sumrule $S_G(3.5)=\int_0^1(dx/x) [F_2^p(x,Q^2)-F_2^n(x,Q^2)]=0.251$. This
is close to the recent value $S_G(4.0)=0.256\pm 0.026$ \cite{arn}, because
of nearly identical contributions from the dominant small $x$ region.
We note in passing that a finite outcome requires $C(0,Q^2)=1$. For CTEQ 
parametrizations from global parton-distribution functions 
$C(0,Q^2)=(1+\delta C(0))$ \cite{lai}, leading to a diverging $S_G$.}.

We now argue that for our purposes the regarded
$Q^2\approx 3.5\, {\rm GeV}^2$ is not $'$large$'$. First we note that
$F_2^{p,n}(x,Q^2)$ has a first inelastic threshold $N+m$, $m$ being the pion
mass, or  $x_{thr}(Q^2)=[1+2Mm/Q^2]^{-1}$, i.e. a finite $x$-distance from 
the elastic point $x=1$. That interval shrinks with increasing $Q^2$, but  
for the case at hand $x_{thr}(3.5)\approx 0.93$, which is marked by a vertical 
line in Fig. 1. Data show that  $F_2^p(x,3.5)$ is negligibly small beyond 
$x\approx 0.9$ \cite{bod1,nicu1,nicu2}.

The above implies that $C(x \to 1,3.5)$ 
can be ascribed to the $NE$ part $F^{N,NE}$, i.e. to static $N$ form factors 
(cf. Eq. (\ref{a13b})). Disregarding the electric form factor  $G_E^n$ of 
the $n$, one finds ($\mu_N$ are the  magnetic moments of the  nucleons)
\begin{eqnarray}
{\lim_{x \to 1}C(x,Q^2)}=\bigg [
\frac{\mu_n\alpha_n(Q^2)}{\mu_p\alpha_p(Q^2)} \bigg ]
\bigg [1+\frac{4M^2}{Q^2}\bigg (\frac {\gamma(Q^2)}{\mu_p}\bigg )^2 
\bigg]^{-1}
\label{a20}
\end{eqnarray}
with
\begin{eqnarray}
\gamma(Q^2)=\frac{\mu_p G_{E}^p(Q^2)}{G_{M}^p(Q^2)}\,;\,
\frac {\alpha_n(Q^2)}{\alpha_p(Q^2)}&=&\frac {G_{M}^n(Q^2)/\mu_n}
{G_M^p(Q^2)/\mu_p}
\label{a21}
\end{eqnarray}
Recent data show that $\gamma,\alpha_p,\alpha_n$ deviate from 1 and equal
for $Q^2=3.5, 5.0\, {\rm GeV}^2$: $\gamma$=0.552, 0.349 and 
$\alpha_n/\alpha_p\approx 1.2, 1.1$ \cite{mjon,sill}. Eqs. (\ref{a20}), 
(\ref{a21}) then yield  
\begin{eqnarray}
C(x=1,3.5)\approx 0.61 \,;\, C(x=1,5.0)\approx 0.56
\label{a22}
\end{eqnarray} 
The large $Q^2$ limit essentially depends on the same for 
$\alpha_n/\alpha_p$. For a value 1, $C(x \to 1,Q^2\to \infty)=0.469$, which
shows that in the above sense $C(1,3.5)$ is still far from a scaling limit.

The above value 0.61 is reasonably close to the extracted one and
has also been entered on the abscissa in Fig.1. Either one definitely 
exceeds previously 
cited values $C(1)\approx 0.42$ from D data (cf. Ref.\onlinecite{meln}).
The above seems to indicate that SU(6) symmetry $u_v(x,3.5)=2d_v(x,3.5)$ 
for the up and down quark parton distribution functions is only mildly broken. 
It is moreover in agreement with globally extracted distribution functions
\cite{lai,bar}.

The above reasoning rests on forms for the nucleon SF, averageKd over 
resonances. Retaining those in detail, the sharp first inelastic threshold 
will become  blurred, but it seems reasonable that $C(1,Q^2)$ for relatively 
low $Q^2$ remains determined by static elastic form factors. That point 
is presently under study. A similar large-$Q^2$ result has been found from 
quark-hadron duality considerations, pushed to the extreme for the  
part $F_2^{N(NE)}$ \cite{meln1}.

Summarizing, we have described methods to extract the neutron Structure 
Function from existing inclusive scattering data on various targets. The 
experimental material from several targets provided  consistent values for 
$C(1,3.5)$. It is very desirable to plan the forthcoming inclusive 
scattering experiments on $D$ and $^4$He with the available 6 GeV beam 
\cite{arrlinu}, such that there be coverage of continuous $x\lesssim 1$-range 
for at least one $Q^2$. Confrontation of those data with accurate
calculations, possible for those nuclei \cite{rtd,viv} will sharpen the 
present outcome for $F_2^n(x,Q^2),C(x,Q^2)$. 

\bigskip

The authors thank G. Petratos for giving detailed information on the proposed 
$A$=3 experiment and J. Arrington for putting at our disposal $D$ and 
Coulomb-corrected NE3 data for C and Fe. ASR profited from discussions with 
G. Salm$\grave{\rm e}$, E. Pace,  several experimentalists at 
JLab and in particular with W. Melnitchuk .

Figures.

Fig. 1.  The ratio $C(x,3.5)=F_2^n(x,3.5)/F_2^p(x,3.5)$ for
$Q=3.5\,$ GeV$^2$  from data on D, C, Fe. The drawn line corresponds to 
$C(1)=0.54$ and the band represents the spread in the result, from averaging 
over different targets and methods. The vertical line for $x=0.93$ marks the 
pion threshold $x_{thr}(3.5)$. The numbers on the right abscissa are quark 
model and QCD predictions for $C(1)$, while 0.61 is the  $NE$ limit 
(\ref{a21}). Also entered are $F_2^p$ and $F_2^n$, corresponding to $C$. 
The band for the latter is hardly noticible in $F_2^n$. 

\begin{table}
\caption{ Values of $C(1,3.5)$ from minimizing variances $w(d_0)$. Results 
are from data on D,C,Fe, using extraction methods I),II). The spread in
results for I correspond to varying $u$-intervals in the variances. No entry 
corresponds to cases for which there is no minimum within the above intervals.}

\begin{tabular}{c||c|c|c|}
   &{D}&{C}&{Fe}\\ \hline
       I   & $0.55\pm0.05$ &   -  & $0.55\pm0.03$ \\ \hline
       II  & 0.55         & 0.50 & -             \\
\end{tabular}
\end{table}


\begin{references}


\bibitem{petr}
G.G. Petratos, Draft version PAC18, Jefferson Lab, July 2000.

\bibitem{afn}
I.R. Afnan, F. Bissey, J. Gomez, A.T. Katramatou, W. Menitchouk,
G.G. Petratos and A.W. Thomas, Phys. Lett. B 493 (2000) 36; F. Bissey,
A.W. Thomas and I.R. Afnan, Phys. Rev. C 64 (2001) 024004.

\bibitem{saito}
K. Saito, C. Boros, K. Tsushima, F. Bissey, I.R. Afnan and A.W. Thomas,
Phys. Lett. B 493 (2000) 288.

\bibitem {pace}
E. Pace, G. Salm$\grave{\rm e}$, S. Scopetta and A. Kievsky, Phys. Rev.
C64 (2000)05523; Nucl. Phys. A 689 (2001) 453. 

\bibitem{sss}
M.M. Sargassian, S. Simula and M.I. Strickman, nucl-th/0105052

\bibitem{aku}
S.V. Akulinichev, S.A. Kulagin and V.M. Vagradov, Phys. Lett. 158 B
(1985) 485; G.V. Dunne, Nucl. Phys. A455 (1986) 701.

\bibitem{gr1}
S.A. Gurvitz and A.S. Rinat, TR-PR-93-77/ WIS-93/97/Oct-PH; Progress in
Nuclear and Particle Physics, Vol. 34 (1995) 245.

\bibitem{atw}
G.B. West, Ann. of Phys. (NY) 74, (1972) 464; W.B. Atwood and G.B. West,
Phys. Rev. D 7, (1973) 773.

\bibitem{pion}
C.H. Llewelyn Smith, Phys. Lett B 128 (1983) 107; M. Ericson and A.W.
Thomas, $ibid$ 112.

\bibitem{rt3}
A.S. Rinat, Proceedings of Meeting 'Prospects on
Hadron and Nuclear Physics', Trieste, IT (1999) World Scientific.

\bibitem{commar}
A.S. Rinat and M.F. Taragin, Phys. Rev. C 62 (2000) 034602.

\bibitem{bod1}   
A. Bodek $et\,al$, Phys. Rev. D 20 (1979) 1471.

\bibitem{liut}
S. Liuti and Franz Gross, Report TH-95-06

\bibitem{umni}  
A.Yu. Umnikov, F.C. Khanna and L.P. Kaptari, Z. F. Physik A   
348 (1994) 211.   

\bibitem{meln} 
W. Melnitchouk and A.W. Thomas, Phys. Lett. B 377 (1996) 11.
 
\bibitem{akul}
S.V. Akulinitchev, S.A. Kulagin and G.M. Vagradov, Phys. Lett. B158 (1985)
485.

\bibitem{amad}   
P. Amadrauz $et\, al$, Phys. Lett B295 (1992) 159; M. Arneodo $et\, al,\,
ibid$ B364 (1995) 107.

\bibitem{bod}    
A. Bodek and J. Ritchie, Phys. Rev. D 23 (1981) 1070.    

\bibitem{cio}
See for instance: C. Ciofi degli Atti, E.Pace and G. Salm$\grave{\rm e}$,
Phys. Rev.  C 43 (1991) 11275; C. Ciofi degli Atti, D.B. Day and
S. Liuti, $ibid$ C 46 (1994) 1045.
 
\bibitem {om1}
O. Benhar, A. Fabrocini, S. Fantoni, G.A. Miller, V.R. Pandharipande and   
I. Sick,  Phys. Rev.  C44 (1991) 2328; Phys. Lett. B359 (1995) 8.

\bibitem{rt1}
A.S. Rinat and M.F. Taragin, Nucl. Phys. A598 (1996); 349 $ibid$ A620,
412 (1997); Erratum $ibid$ A623 (1997) 773.

\bibitem{gr2}  
S.A. Gurvitz and A.S. Rinat, nucl-th/0106032; Phys. Rev. C 65, to be
published.

\bibitem{rt11}
A.S. Rinat and M.F. Taragin, Phys Rev. C 60 (1999) 044601.

\bibitem{arrold}
J.Arrington $et\,al$, Phys. Rev. C 53 (1996) 224

\bibitem{gomez}
J. Gomez $et\,al$, Phys. Rev. D 49 (1994) 4348.

\bibitem{bench}
H. Kamada $et\,al$, Phys. Rev. C 64 (2001) 044001.

\bibitem{rtd}
A.S. Rinat and M.F. Taragin, Phys. Rev. C 65 (2002) 041610.

\bibitem{viv}
M. Viviani, A. Kievsky and A.S. Rinat, nucl-th/0111048, submitted to Phys.
Rev. C

\bibitem{nicu1}
I. Niculescu $et\,al$, Phys. Rev. Lett. 85 (2000) 1182.

\bibitem{arrd}  
J. Arrington, private communication.

\bibitem{arr1}
J. Arrington $et\, al$, Phys. Rev. Lett. 82 (1999) 2056; CalTech
PhD thesis 1998.

\bibitem{ne3}
D.B. Day $et\,al$, Phys. Rev. C 40 (1993) 1849.

\bibitem{dasu}
S. Dasu $et\,al$ Phys. Rev. Lett. 61 (1988), 1061; Phys. Rev. D 49 (1994)
5641; P.E. Bosted $et\,al$, Phys. Rev. C 46 (1992) 2505;
B.W. Fillipone $et\,al$, Phys. Rev. C 45 (1992) 1582.

\bibitem{nicu2}
I. Niculescu $et\,al$, Phys. Rev. Lett. 85 (2000) 1186.

\bibitem{mjon}
M. Jones $et\,al$, Phys. Rev. Lett. 84 (2000) 1398; Third Workshop on
'Perspective in Hadronic Physics' Trieste 2001, IT; to be published.

\bibitem{sill}
A.F Sill $et\,al$, Phys. Rev. D 48 (1993) 29.

\bibitem{arn}    
M. Arneodo $et\,al$, Phys. Rev. D 50 (1994) R1.

\bibitem{lai}
H.L. Lai $et\,al$, Phys. Rev. D 51 (1995) 4763; $ibid$ D 55 (1997) 1280;
Eur. Phys. J. C 12 (2000) 375.

\bibitem{bar}
V. Barone, C. Pascaud and F. Zomer, Eur. Phys. J. C 12 (2000) 243,

\bibitem{meln1}
W. Melnitchouk, Phys. Rev. Lett. 86 (2001) 35.

\bibitem{arrlinu}
J. Arrington $et\,al$, Spokesman proposal JLab, May 2000.

\end{references}
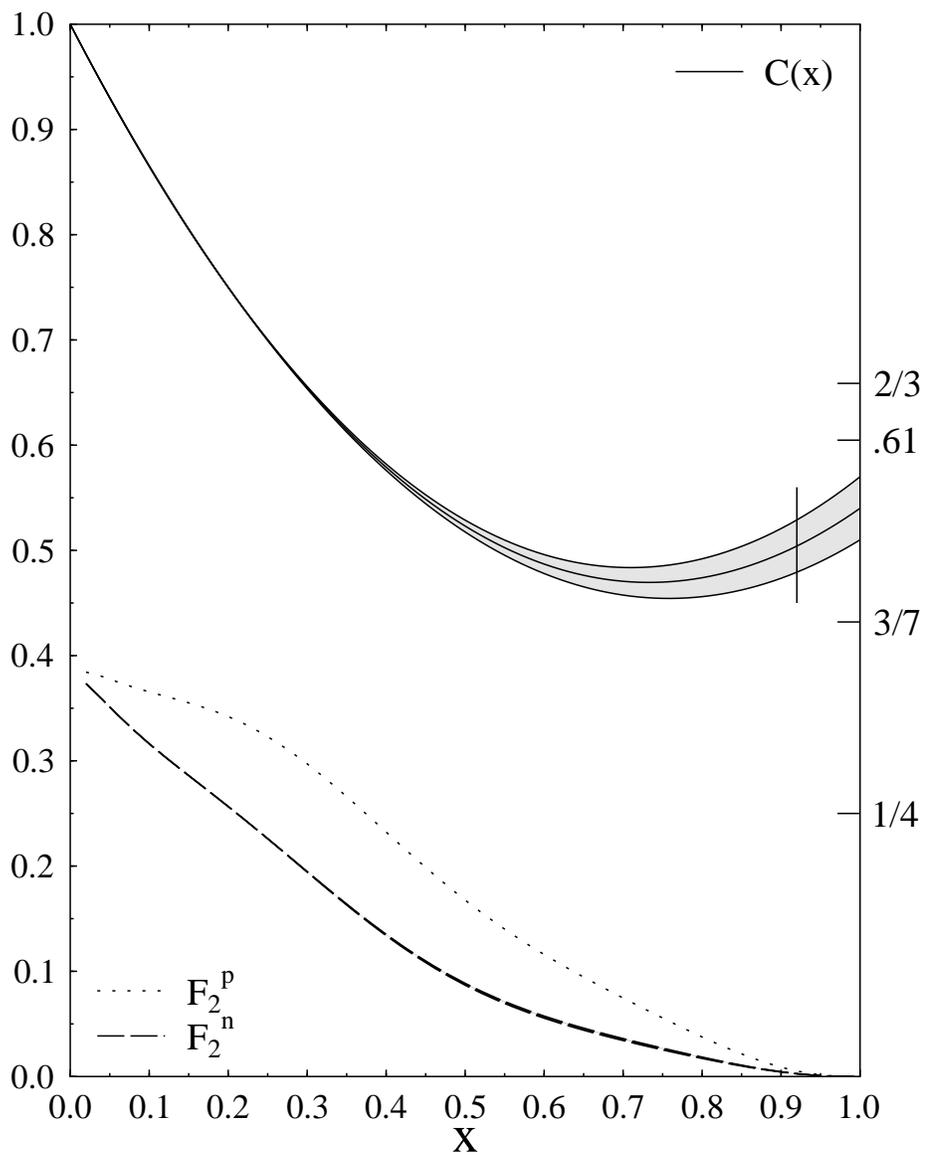
\end{document}